\documentclass[12pt]{iopart}
\eqnobysec

\usepackage{hyperref}
\usepackage[]{cleveref}
\usepackage{amssymb}
\usepackage{graphicx}
\usepackage{iopams}

\newcommand{\be}{\begin{equation}}
\newcommand{\ee}{\end{equation}}
\newcommand{\nn}{\nonumber}

\newcommand{\lp}{\left(}
\newcommand{\rp}{\right)}
\newcommand{\lbk}{\left[}
\newcommand{\rbk}{\right]}

\newcommand{\di}{\mathrm{d}}
\newcommand{\Cov}{\mathrm{Cov}}
\newcommand{\fent}{F_{\rm{ent}}}

\newcommand{\xbf}{\bi{x}} %

\begin{document}

\title{Branching processes with resetting as a model for cell division}

\author{Arthur Genthon$^1$, Reinaldo Garc\'{i}a-Garc\'{i}a$^2$, David Lacoste$^1$}

\address{$^1$ Gulliver UMR CNRS 7083, ESPCI Paris, Universit\'{e} PSL, 75005 Paris, France}
\address{$^2$ Departamento de F\'{i}sica y Matem\'{a}tica Aplicada, Facultad de Ciencias, Universidad de Navarra, c/Irunlarrea 1, E-31008, Pamplona, Navarra, Spain}

\ead{arthur.genthon@espci.fr}

\vspace{10pt}
\begin{indented}
\item[] \today
\end{indented}

\begin{abstract}
We study the Stochastic Thermodynamics of cell growth and division using a theoretical framework based on branching processes with resetting. Cell division may be split into two sub-processes: branching, by which a given cell gives birth to an identical copy of itself, and resetting, by which some properties of the daughter cells (such as their size or age) are reset to new values following division. We derive the first and second laws of Stochastic Thermodynamics for this process, and identify separate contributions due to branching and resetting. 
We apply our framework to well-known models of cell size control, such as the sizer, the timer, and the adder. We show that the entropy production of resetting is negative and that of branching is positive for these models in the regime of exponential growth of the colony. This property suggests an analogy between our model for cell growth and division and heat engines, and the introduction of a thermodynamic efficiency, which quantifies the conversion of one form of entropy production to another.
\end{abstract}

\vspace{2pc}

\section{Introduction}

Resetting, a stochastic process involving an instantaneous transition to a pre-defined position or region of space, has been a very active field of research for the past ten years, triggered by the seminal paper by Evans and Majumdar in 2011 \cite{evans_diffusion_2011}. Various groups have extensively studied such processes, in the context of target search \cite{coppey_kinetics_2004,benichou_optimal_2005}, or in the context of non-equilibrium steady-states (NESS) and first passage times \cite{evans_stochastic_2020}. 
The first study of resetting from the point of view of stochastic thermodynamics was carried out in 2016, by Fuchs et al. \cite{fuchs_stochastic_2016} who derived the first and second laws of thermodynamics for resetting to a single fixed position $x_0$. Later, Rold{\'a}n et al. 
developed a path integral approach for resetting \cite{roldan_path-integral_2017}, 
which was used by Pal et al. to derive integral fluctuation relations \cite{pal_integral_2017,gupta_work_2020} and thermodynamic uncertainty relations \cite{pal_thermodynamic_2021} for systems with resetting. 
To our knowledge, the combination of stochastic resetting and branching processes has only been considered for the purpose of search strategies \cite{eliazar_branching_2017,pal_first_2019}. In particular, the authors compared first passage times for branching search, a term coined by Eliazar \cite{eliazar_branching_2017} to describe a strategy mixing branching and resetting, and for resetting alone. However, the thermodynamics of stochastic processes involving both resetting and branching at the same time has not been studied so far. 

Branching processes with resetting are not only useful in the context of target search but are fundamental in biology, in particular to describe populations of cells that grow and divide. Indeed, cell division intrinsically features branching, since one mother cell gives birth to two daughters cells. As a result of division, some traits of daughter cells are reset to an absolute value, such as the age of the cell which is reset at $0$, while others are reset to a value relative to that of the mother at the moment of the division. For instance, the size of the daughter cell restarts at a value close to half the size of the mother cell for organisms undergoing symmetric division. In order to study cell division from the point of view of stochastic thermodynamics, we extend the analysis of Fuchs et al. \cite{fuchs_stochastic_2016} to allow a restart at a relative position instead of an absolute one, and to incorporate branching in the formalism.

In \cref{sec_thermo}, we present our model describing a population of overdamped Brownian particles undergoing relative resetting and branching in a 1-dimensional potential. We then derive the first and second laws of thermodynamics for this model and identify the separate contributions of resetting and branching. Next, we propose an alternative version of the second law, which applies to athermal systems, and we show how our results generalize to the case of an $n$-dimensional problem.
In \cref{sec_cells}, we explain why cell division can be understood as a combination of branching and resetting processes. Then, focusing on some of the most common cell size control mechanisms, we obtain analytical expressions for the work and entropy terms associated with resetting and branching. 
Stochastic thermodynamics has been often used to analyze information processing and efficiency of small biological systems. In the same spirit, we propose an analogy between cell division and stochastic heat engines, which leads us to introduce an efficiency, which quantifies a form of conversion of entropy production from resetting to branching. We study this efficiency for known models of cell size control.

\section{Thermodynamics of branching processes with stochastic resetting to a relative position in continuous space}
\label{sec_thermo}

\subsection{Model}

\begin{figure}
	\centering
	\includegraphics[width=0.7\linewidth]{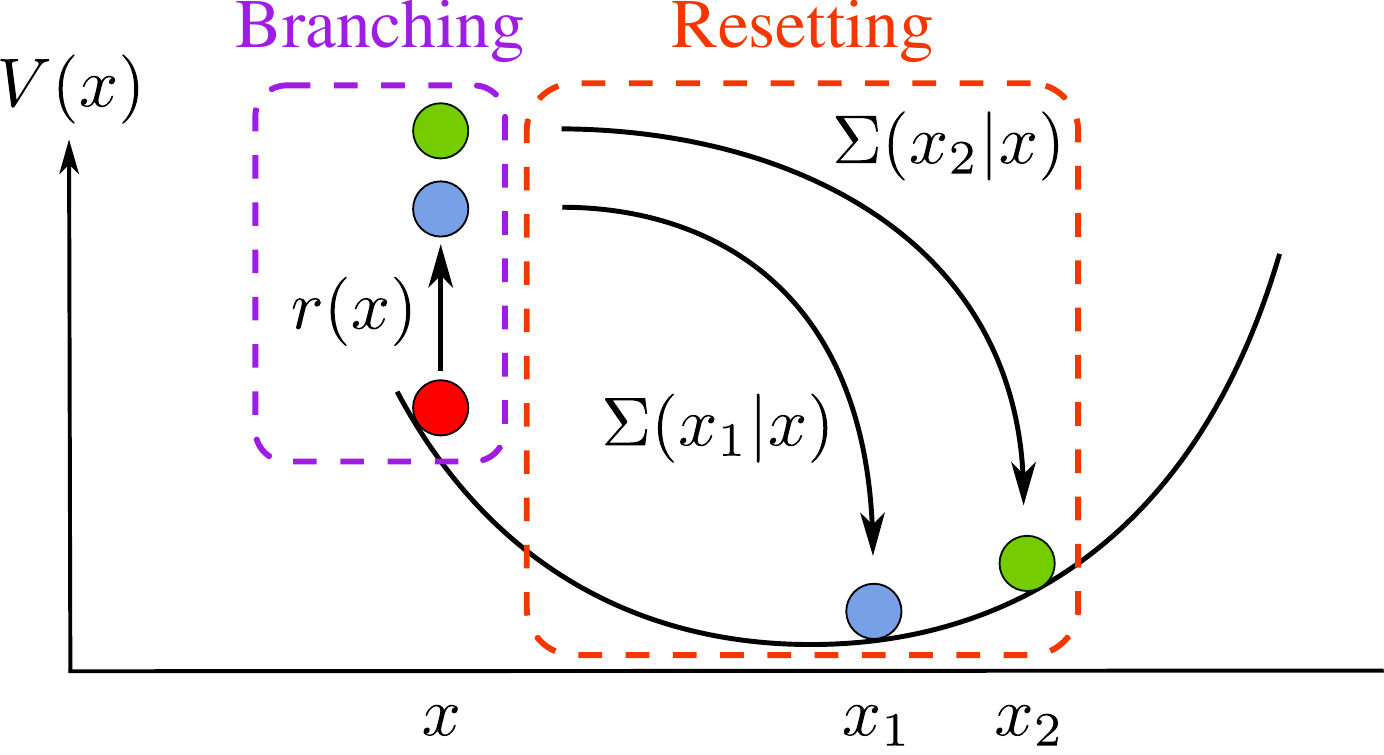}
	\caption{Illustration of a particle branching into $m=2$ particles, each of which is independently reset to a position relative to the branching position $x$ through the transition probability $\Sigma$. The parabolic potential shown here is for illustration only, potentials can be arbitrary in our model.}
	\label{fig_rst_brc}
\end{figure}

We consider a population of overdamped Brownian particles with mobility $\mu$ and diffusion coefficient $D$, in contact with a heat bath at temperature $T$ and subject to a potential $V$. For simplicity, we treat the case of a $1$-dimensional potential $V(x)$ in the main text and show in \ref{sec_nd} that our results hold in $n$ dimensions. In the following we assume Einstein relation $D= \mu T$, with unit Boltzmann's constant $k_B=1$. 
In addition to their diffusive motion, particles randomly branch at a space-dependent rate $r(x)$, leading one particle to give birth to $m$ particles (including the original one) at the same position. Instantly after branching, all $m$ particles are reset to new positions with the transition probability $\Sigma(x|x')$, for a restart at position $x$ if the original particle branched at position $x'$, as illustrated on \cref{fig_rst_brc}. Since the positions of the new particles depend on $x'$, we call it \textit{relative} resetting, as opposed to \textit{absolute} resetting, where particles either restart at a fixed position $x_0$ ($\Sigma(x|x')=\delta(x-x_0)$) or restart at a random position with a probability distribution $\Sigma(x|x')=f(x)$, independently of $x'$.
In this case, an explicit solution of the NESS is in general no longer available, unlike what happens for  absolute resetting. 
The dynamics of the number $n(x)$ of particles at position $x$ at time $t$ is described by the following generalized Fokker-Planck equation:
\be
\label{eq_FP_n_1d}
\fl
\partial_t n(x) = -\partial_x \lbk \mu F(x) n(x) - D \partial_x n(x) \rbk - r(x) n(x) + m \int \di x' \Sigma(x|x') r(x') n(x') \,,
\ee
where $F(x)= - \partial_x V$ is the conservative force deriving from the potential $V$.
We recast this equation at the probability level by defining the proportion of particles at position $x$ at time $t$: $p(x)=n(x)/N_t$, with $N_t=\int \di x \ n(x,t)$ the total number of particles at time $t$:
\be
\label{eq_FP_1d}
\partial_t p(x) = -\partial_x j(x) - \lbk \Lambda + r(x) \rbk p(x) + m \int \di x' \Sigma(x|x') r(x') p(x') \,,
\ee
where we have defined the current $j(x) = \mu F(x) p(x) - D \partial_x p(x)$ and the instantaneous population growth rate $\Lambda=(\partial_t N_t)/N_t$. We take vanishing boundary conditions for probability $p$ and current $j$ at $x=0$ and $x \rightarrow +\infty$.

\subsection{First and second laws of thermodynamics}
\label{sec_1_2_laws}

We follow the approach from \cite{fuchs_stochastic_2016} to derive the first and second laws of thermodynamics from \cref{eq_FP_1d}. First, we multiply \cref{eq_FP_1d} by the potential $V(x)$ and integrate over $x$ to obtain the time evolution of the internal energy $U=\int \di x \ V(x) p(x)$:
\begin{eqnarray}
\fl
\label{eq_1_law}
-\dot{U}= \int \di x \ j(x) F(x) +  m \int \di x \ r(x) p(x) \lbk V(x) - \langle V \rangle_{\rho_{nb}} \rbk  \nn \\
+\Lambda U - (m-1) \int \di x \ r(x) p(x) V(x) \,,
\end{eqnarray}
where we have introduced the newborn position distribution
\be
\rho_{nb}(x)=\frac{\int \di x' \ \Sigma(x|x') r(x') p(x')}{\int \di x' \ r(x') p(x')} \,,
\ee
defined as the ratio of the rate of birth of new particles at position $x$ to the rate of birth of the total number of newborn particles. 
The notation $ \langle \cdot \rangle_{\rho_{nb}}$ indicates the average value with respect to the distribution $\rho_{nb}$. By default, averages values, variances and covariances are implicitly computed with the probability distribution $p$.

Following \cite{seifert_stochastic_2012,fuchs_stochastic_2016}, we identify the first term as the rate at which heat is transferred from the system to the thermostat:
\be
\label{eq_Q}
\dot{Q}=\int \di x \ j(x) F(x) \,.
\ee
The second term is the rate at which work is extracted from the system due to the resetting of particles to their new positions:
\be
\label{eq_W_rst}
\dot{W}_{\rm{rst}}=m \int \di x \ r(x) p(x) \lbk V(x) - \langle V \rangle_{\rho_{nb}} \rbk \,.
\ee
The last two terms are interpreted as the work extraction rate from the system due the branching of particles. They can be written more explicitly by using the following expression of the population growth rate, obtained by integrating \cref{eq_FP_1d} over $x$:
\be
\Lambda=(m-1) \int \di x \ r(x) p(x) \,.
\label{eq_Lb}
\ee
Then
\begin{eqnarray}
\label{eq_W_brc_sum}
\dot{W}_{\rm{brc}}&=\Lambda U - (m-1) \int \di x \ r(x) p(x) V(x) \\
\label{eq_W_brc_cov}
&=-(m-1)\Cov(r,V) \,.
\end{eqnarray}
This contribution is null if $r(x)$ and $V(x)$ are independent, which is trivially the case when at least one of the two functions is constant. Indeed, the average internal energy is not affected by the apparition of newborn particles at any position if the energy landscape $V(x)$ is flat; nor if the branching rate is constant, so that branching affects equally all particles regardless of their positions and thus has no impact on $p(x)$. 

Finally, the first law for branching processes with relative resetting reads:
\be
\label{eq_1st_law}
-\dot{U}=\dot{Q}+\dot{W}_{\rm{rst}}+\dot{W}_{\rm{brc}} \,,
\ee
where we count positively work extracted from the system and heat dissipated into the environment. 

The Shannon entropy of the system is defined by
\be
S_{\rm{sys}}= -\int \di x \ p(x) \ln p(x) \,.
\ee
To derive the second law, we take the time derivative of this entropy and use \cref{eq_FP_1d}, which gives:
\begin{eqnarray}
\fl \dot{S}_{\rm{sys}} = - \int \di x \ \frac{j(x) \partial_x p(x)}{p(x)}  - \Lambda S_{\rm{sys}} - (m-1) \int \di x \ r(x) p(x) \ln p(x) \nn \\
 + m \int \di x \ r(x) p(x) \lbk \ln p(x) - \langle \ln p \rangle_{\rho_{nb}} \rbk \,,
\end{eqnarray}
where the term due to the current $j(x)$ can be split into two contributions if the temperature is non-zero (the case $T=0$ is discussed in \cref{sec_T_0}):
\be
\label{eq_decomp_curr}
- \int \di x \ \frac{j(x) \partial_x p(x)}{p(x)}=\int \di x \ \frac{j^2(x)}{D p(x)} - \int \di x \ \frac{\mu j(x) F(x)}{D} \,.
\ee

We identify four contributions to the rate of change in the entropy of the system, respectively the entropy production rate due to the heat exchange with the thermostat $\dot{S}_{\rm{m}}$, the entropy production rate of non-equilibrium current $j(x)$: $\dot{S}_{\rm{c}}$, the branching entropy production rate $\dot{S}_{\rm{brc}}$ and the resetting entropy production rate $\dot{S}_{\rm{rst}}$:
\be
\label{eq_S_m}
\dot{S}_{\rm{m}} = \frac{\dot{Q}}{T} = \int \di x \ \frac{\mu j(x) F(x)}{D} \,,
\ee
where we used Einstein's relation $D=\mu T$,
\be
\dot{S}_{\rm{c}}=\int \di x \ \frac{j^2(x)}{D p(x)} \geq 0
\ee
\begin{eqnarray}
\dot{S}_{\rm{brc}} &=- \Lambda S_{\rm{sys}} - (m-1) \int \di x \ r(x) p(x) \ln p(x) \\
\label{eq_S_brc_cov}
&= -(m-1) \Cov (r, \ln p)
\end{eqnarray}
\be
\label{eq_S_rst}
\dot{S}_{\rm{rst}}=m \int \di x \ r(x) p(x) \lbk \ln p(x) - \langle \ln p \rangle_{\rho_{nb}} \rbk
\ee
Using the positivity of the entropy production rate of non-equilibrium currents, the second law for branching processes with relative resetting in steady state ($\dot{S}_{\rm{sys}}=0$) reads:

\be
\label{eq_2law_ineq_v1}
\dot{S}_{\rm{rst}}+\dot{S}_{\rm{brc}} \leq \dot{S}_{\rm{m}}
\ee

The first and second laws we derived reduce to the ones obtained in \cite{fuchs_stochastic_2016} if there is no branching and if the particle is reset to a fixed position $x_0$. Indeed, setting $m=1$ and thus $\Lambda=0$, leads to $\dot{W}_{\rm{brc}}=0$ and $\dot{S}_{\rm{brc}}=0$, and therefore \cref{eq_1st_law} and \cref{eq_2law_ineq_v1} read respectively $-\dot{U}=\dot{Q}+\dot{W}_{\rm{rst}}$ and $\dot{S}_{\rm{rst}} \leq \dot{S}_{\rm{m}}$. In our framework, absolute resetting to fixed position $x_0$ is obtained by setting $\Sigma(x|x')=\delta(x-x_0)$, then $\rho_{nb}(x)=\delta(x-x_0)$ and thus $\langle V \rangle_{\rho_{nb}} = V(x_0)$ in \cref{eq_W_rst} and $\langle \ln p \rangle_{\rho_{nb}} = \ln p(x_0)$ in \cref{eq_S_rst}, in agreement with \cite{fuchs_stochastic_2016}.

Instead, when there is branching but no resetting, i.e. when particles randomly multiply and then continue to diffuse from the same position, the transition probability is given by $\Sigma(x|x')=\delta(x-x')$, and thus $\rho_{nb}(x)=r(x) p(x)/\int \di x' \ r(x') p(x')$. In this case, $\int \di x \ r(x) p(x) V(x) =  \int \di x \ r(x) p(x) \langle V \rangle_{\rho_{nb}}$ in \cref{eq_W_rst}, leading to $\dot{W}_{\rm{rst}}=0$, and similarly $\dot{S}_{\rm{rst}}=0$. 
Without resetting, the first and second laws finally read $-\dot{U} = \dot{Q} +\dot{W}_{\rm{brc}}$ and $\dot{S}_{\rm{brc}} \leq \dot{S}_{\rm{m}}$ respectively.

\subsection{Alternative form of the second law}
\label{sec_alt_2_law}

To derive the second law in the previous section, we decomposed \cref{eq_decomp_curr} into two terms: a rate of change in medium entropy 
$\dot{S}_{\rm{m}}$ due to the heat exchange with surrounding heat bath, and a positive entropy production rate
$\dot{S}_{\rm{c}}$ due to non-equilibrium currents. 

Alternatively, another decomposition is obtained by replacing the current $j(x)$ by its definition: 
\be
\label{eq_decomp_curr_2}
- \int \di x \ \frac{j(x) \partial_x p(x)}{p(x)}=\mu \int \di x \ p(x) \partial_x F(x) + \mu T \int \di x \ p(x) \lp \partial_x \ln p(x) \rp^2 \,,
\ee
which leads to:
\be
\label{eq_2nd_law_v2}
\dot{S}_{\rm{sys}}=\dot{S}_{\rm{brc}}+\dot{S}_{\rm{rst}}+ \dot{S}_{\rm{fd}} + \mu T \int \di x \ p(x) \lp \partial_x \ln p(x) \rp^2 \,,
\ee
where we introduced the average force divergence $\dot{S}_{\rm{fd}}= \mu \langle \partial_x F \rangle$. To establish a closer connection with the discussion of the athermal case (see next section), it is useful to combine the last term in~\cref{eq_2nd_law_v2} with $\dot S_{\rm{fd}}$ by using the notion of entropic force, $\fent(x)=-T\partial_x \ln p(x)$. With this, one can introduce a generalized force $\tilde{F}(x)=F(x)+\fent(x)\equiv F(x)-T\partial_x \ln p(x)$, and the corresponding generalized average force divergence contribution to the entropy production rate, $\dot{\tilde{S}}_{\rm{fd}}=\mu\langle\partial_x\tilde{F}\rangle$. Note that we have:
\begin{eqnarray}
\label{entropic-force-gradient}
\mu T \int \di x \ p(x) \lp \partial_x \ln p(x) \rp^2 &=-\mu\int \di x \ p(x) \partial_x \ln p(x)\ \fent(x)\nonumber\\
&=-\mu\int \di x \ \partial_xp(x) \ \fent(x)\nonumber\\
&=\mu\int \di x \ p(x)\partial_x\fent(x)\nonumber\\
&\equiv\mu\langle\partial_x\fent\rangle, 
\end{eqnarray}
which leads to the result
\be
\label{entropic-force-gradient-1}
\dot{S}_{\rm{fd}} + \mu T \int \di x \ p(x) \lp \partial_x \ln p(x) \rp^2=\mu\langle\partial_x[F+\fent]\rangle=\dot{\tilde{S}}_{\rm{fd}}.
\ee
We then rewrite~\cref{eq_2nd_law_v2} in the following equivalent form:
\be
\label{eq_2nd_law_v2-1}
\dot{S}_{\rm{sys}}=\dot{S}_{\rm{brc}}+\dot{S}_{\rm{rst}}+ \dot{\tilde{S}}_{\rm{fd}}.
\ee

Using the positivity of the last term in the r.h.s. of \cref{eq_2nd_law_v2}, we obtain an alternative version of the second law in steady state:
\be
\label{eq_2nd_law_ineq_v2}
\dot{S}_{\rm{brc}}+\dot{S}_{\rm{rst}} \leq - \dot{S}_{\rm{fd}} \,.
\ee

The two versions of the second law \cref{eq_2law_ineq_v1,eq_2nd_law_ineq_v2} provide different bounds for the entropy production rate due to branching and resetting, so that by combining them we have:
\be
\label{eq_2nd_law_ineq_v3}
\dot{S}_{\rm{brc}}+\dot{S}_{\rm{rst}} \leq \min \lp - \dot{S}_{\rm{fd}}, \dot{S}_{\rm{m}} \rp \,.
\ee
In the limit of large temperatures, the current $j(x)=\mu F(x) p(x) - D \partial_x p(x)$ becomes 
purely diffusive $j(x) \simeq - D \partial_x p(x)$, and as a result, the medium entropy
$\dot{S}_{\rm{m}}=\mu \langle \partial_x F \rangle =\dot{S}_{\rm{fd}}$. 
Thus, in this limit the upper bound reads $\dot{S}_{\rm{brc}}+\dot{S}_{\rm{rst}} \leq - | \dot{S}_{\rm{fd}} |.$ The behavior in the opposite limit of vanishing temperature is examined in 
more details in the next section.

\subsection{Athermal systems}
\label{sec_T_0}

Until now, we considered particles in contact with a heat bath, which allowed us to use standard stochastic thermodynamics. However, athermal systems where $T=0$ are important to describe situations where particles move deterministically between branching events. In that case, there is no diffusion, only the deterministic force $F(x)$ is present, but branching and resetting events are still stochastic. Such situations are relevant biogically as we discuss in \cref{sec_cells}. 

The first law is still mathematically valid, even if the current $j(x)=\mu F(x)p(x)$ is only convective. We find $\dot{Q}=\mu \langle F^2 \rangle$, and we call the corresponding quantity $Q$ an athermal heat, although it is important to emphasize that this quantity cannot be interpreted as an exchange of heat with a thermostat. 

The second law in the form of \cref{sec_1_2_laws} is no longer valid, since $\dot{S}_{\rm{m}}$ and $\dot{S}_{\rm{c}}$ diverge in the case $T=0$. However, the other decomposition of the Shannon entropy (\cref{sec_alt_2_law}) remains defined because the last term in \cref{eq_2nd_law_v2} (entropic contribution) tends to $0$ as $T \rightarrow 0$, since the integral is not singular in practice. We propose to call the following equality 
\be
\label{eq_2nd_atherm}
\dot{S}_{\rm{sys}}=\dot{S}_{\rm{brc}}+\dot{S}_{\rm{rst}}+\dot{S}_{\rm{fd}} \,,
\ee
the second law for athermal systems (despite the absence of a corresponding inequality) because it corresponds to 
the $T=0$ version (i.e., in absence of entropic forces) of~\cref{eq_2nd_law_v2-1}.

\section{Application to cell size control}
\label{sec_cells}

As stated in the introduction, branching processes with resetting appear in biology, for example in the context of cell division. Indeed, when cells divide they give birth to $m=2$ daughter cells, and many cell properties are affected by division. 
Cell division provides examples of absolute resetting, for instance age which is reset at $0$ for both daughter cells independently of the age of the dividing mother cell, and of relative resetting, such as for the volume or the number of proteins for example, that are split at division between the two daughter cells. 
The idea to consider the cell size jump at division as a form of stochastic resetting was first mentioned in \cite{garcia-garcia_linking_2019}.

In this section, we use the results obtained in \cref{sec_thermo} in the context of growing colonies of cells, focusing in particular on the three most common division strategies \cite{robert_division_2014,jun_fundamental_2018,ho_modeling_2018}: the sizer, the timer and the adder.

\subsection{Sizer}
\label{sec_sizer}

The sizer mechanism assumes that the division rate is a function of the size of the cell only, hence \cref{eq_FP_n_1d} describes the dynamics of a population of cells where the position $x$ of the particle is understood as the cell size. We use the terms size and volume equivalently, which is common in the cell size-control literature since bacteria grow mainly along one direction. The volume partition between the two daughter cells is modeled by $\Sigma(x|x')$, which is centered around $x'/2$ for bacterial cells, since each daughter cell inherits on average half the volume of the dividing cell. More generally, this stochastic partitioning of volume can also be asymmetric, when $\Sigma(x|x')$ has the form of a Gamma distribution, which leads to a beta distribution for the volume ratio $x/x'$ \cite{jia_cell_2021}. Moreover, in a rich medium, bacteria generally follow an exponential growth between divisions with additional noise: 
\be
\label{eq_x_lang}
\dot{x}=\nu  x +\xi \,,
\ee
with $\nu$ the single-cell growth rate, which is non-fluctuating and strictly positive.
This is a Langevin equation with unit mobility $\mu=1$, force $F(x)=\nu x$, and Gaussian noise $\xi$ with variance $2T$. The force $F$ derives from a potential $V(x)=-\nu x^2/2$, which is non-confining, unlike the example shown in \cref{fig_rst_brc}.

First, $V(x)$ is a decreasing function of $x$ while the size regulation imposes that the division rate $r(x)$ be an increasing function of $x$. Thus, the covariance between those two functions is negative, which results in a positive production of work due to branching by eq. (\ref{eq_W_brc_cov}): $\dot{W}_{\rm{brc}} \geq 0$.
In contrast, the decrease of $V(x)$ with $x$ leads to a negative production of work due to resetting. This can be seen easily from eq. (\ref{eq_W_brc_sum}) when replacing the average value computed with the newborn distribution by its definition:
\be
\dot{W}_{\rm{rst}}=2 \int \di x \ r(x) p(x) \lbk V(x) - \int \di x' \  \Sigma(x'|x) V(x') \rbk \,,
\ee
where the term in the bracket is the difference between the potential for a cell of given size $x$ and the average value of the potential for a new cell, born from the division of a cell of size $x$. Due to the non-confining shape of potential $V(x)$, the restarting size $x$ has a higher internal energy than dividing sizes $x'$, thus this difference is negative for any $x$ and $\dot{W}_{\rm{rst}} \leq 0$. 
Note that with the sign convention chosen in the first law \cref{eq_1st_law}, $\dot{W}_{\rm{rst}} \leq 0$ corresponds indeed to an increase of the internal energy of the population of cells because of resetting. 

Second, our model includes the case of noisy cell growth between two divisions; however, the majority of models in the literature focus on deterministic growth, which is obtained by setting $\xi=0$ in the Langevin representation \cref{eq_x_lang} and $D=0$ (or equivalently $T=0$) in the Fokker-Planck representation \cref{eq_FP_1d}. This case is covered by \cref{sec_T_0} on athermal systems and the steady-state second law equality reads
\be
\label{eq_2law_sizer}
\Lambda = -\dot{S}_{\rm{rst}}-\dot{S}_{\rm{brc}} \,,
\ee
where we used that $\dot{S}_{\rm{fd}}=\langle \partial_x F \rangle = \nu$, and 
the property that in a steady-state, the population growth rate $\Lambda$ must be equal to the single cell growth rate $\nu$. This property is derived as follows: one integrates of \cref{eq_FP_1d} after multiplication by size $x$ to obtain $\partial_t \langle x \rangle =(\nu - \Lambda) \langle x \rangle$, and thus $\Lambda=\nu$ in steady-state. Intuitively, the total volume $\mathcal V$ of the population grows at a rate $\nu$ since each cell grows at this rate, so that the mean volume $\langle x \rangle = \mathcal V/N \sim \exp \lbk \lp \nu - \Lambda \rp t \rbk $ converges only if $\Lambda =\nu$ \cite{lin_effects_2017}. Moreover, in this athermal case, the rate of athermal heat introduced in \cref{sec_T_0} is now given by $\dot{Q}=\nu^2$. In the following, we focus on deterministic growth only.

Third, when following a single lineage of cells, for example in a mother machine setup \cite{wang_robust_2010}, then $\dot{S}_{\rm{brc}}=0$ since there is no possibility of branching in that case. As a result, the second law in steady-state (\cref{eq_2law_sizer}) reduces to $\dot{S}_{\rm{rst}}=-\nu$. Interestingly, in this case, the entropy production rate due to resetting depends only on the single cell growth rate, but not on the division rate $r(x)$ nor on the kernel $\Sigma$. 

Finally, the signs of the branching and resetting entropy production rates can be obtained with
reasonable assumptions. 
Experimental steady-state size distributions are well fitted by log-normal functions as long as the transition probability $\Sigma$ is peaked around symmetric division (see \cite{hosoda_origin_2011} for a data collapse of the log size distributions from various datasets on a Gaussian function). Note that log-normal functions are not rigorously solutions of \cref{eq_FP_1d}, but since they are accurate approximations of the true solution, we use them as an ansatz for the steady-state solution.
Moreover, division rates inferred from experimental data show a power law dependence on size for non-extreme sizes \cite{robert_division_2014,osella_concerted_2014}, and the log-normal ansatz computed with a power-law division rate is in good agreement with experiments \cite{hosoda_origin_2011}. The theoretical origin of the power-law dependence of the division rate on size has also been discussed in Ref. \cite{nieto_unification_2020}.
Assuming that $p = \rm{Lognormal} (\mu ,\sigma^{2})$ and $r(x)=x^{\alpha}$, we show in \ref{app_S_brc} that the covariance in eq. (\ref{eq_S_brc_cov}) is computable and given by:
\be
\label{eq_S_brc_sizer}
\dot{S}_{\rm{brc}} =  \alpha \sigma^2 \lp 1 + \frac{\alpha}{2} \rp \exp \lbk \alpha \mu +  \frac{\alpha^2 \mu^2}{2} \rbk \geq 0 \,,
\ee
which is positive regardless of the values of $\alpha \geq 0$, $\mu$ and $\sigma$.
Using this inequality, \cref{eq_2law_sizer} imposes that $\dot{S}_{\rm{rst}} \leq 0$, meaning that resetting is a way to avoid extremely large cells and thus to reduce the Shannon entropy of the size distribution at steady-state. However, this constraint no longer exists in the transient dynamics, where resetting can increase the Shannon entropy of the size distribution: $\dot{S}_{\rm{rst}} \geq 0$.

\begin{figure}
	\centering
	\includegraphics[width=0.7\linewidth]{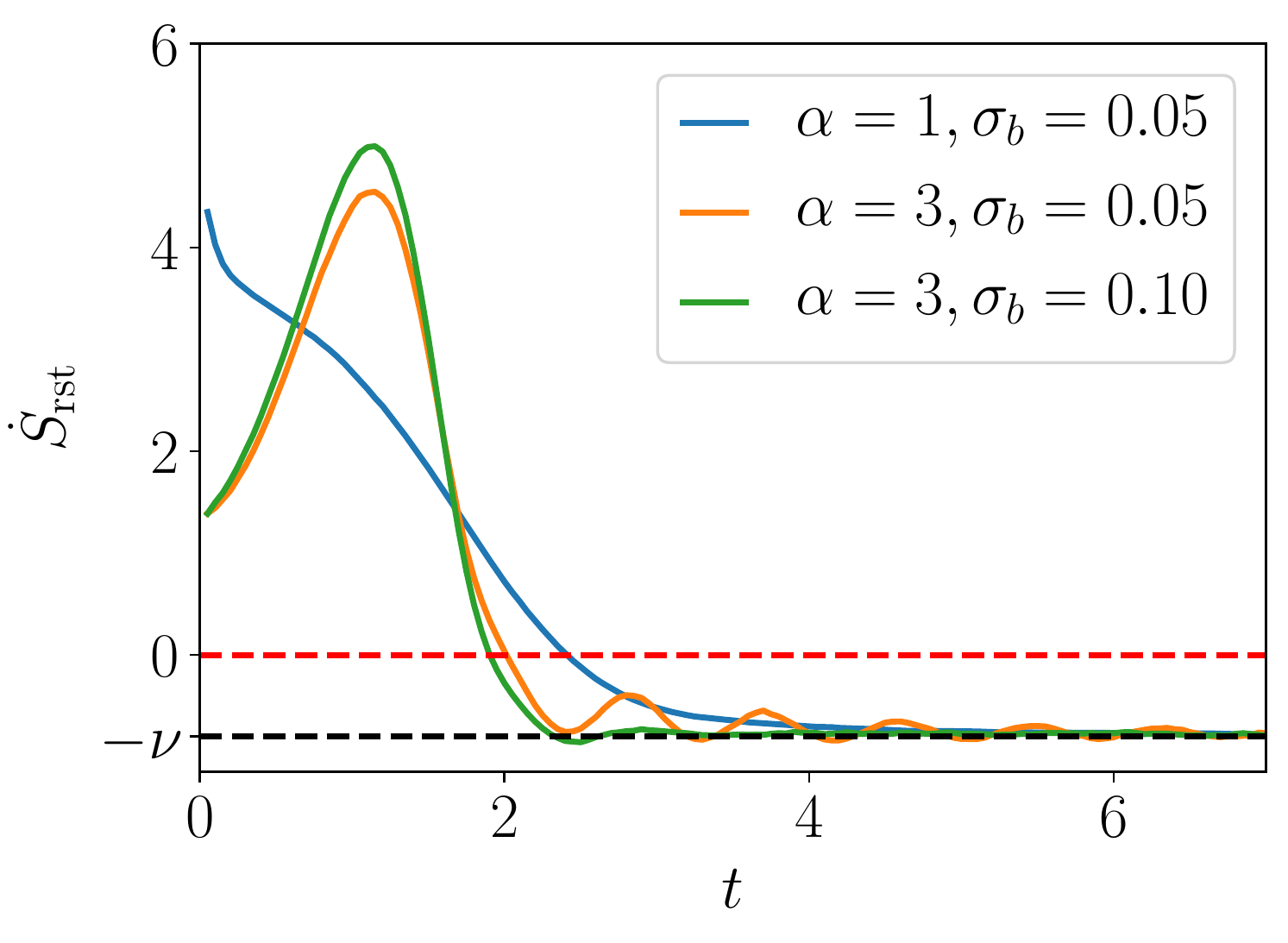}
	\caption{Time evolution of the entropy production rate due to resetting $\dot{S}_{\rm{rst}}$, in the case of a size-control mechanism with rate $r(x)=x^{\alpha}$, deterministic growth between division at a rate $\nu=0.8$ and Gaussian kernel $\Sigma(\cdot|x') = \mathcal{N}(x'/2,(\sigma_{b} x')^2)$ for the partition of volume between the daughter cells at division. Each curve corresponds to a choice of parameters ($\alpha$, $\sigma_{b}$), and is the result of 100000 single lineage simulations (without branching so $\dot{S}_{\rm{brc}}=0$), starting at size $x_0=0.5$. 
	All curves converges to the black dashed line at $\dot{S}_{\rm{rst}}=-\nu$ in the long time limit.}
	\label{fig_s_rst_vs_t}
\end{figure}

To illustrate the last two points, we show in \cref{fig_s_rst_vs_t} the evolution of $\dot{S}_{\rm{rst}}$ with time. We simulated $10^5$ independent single lineages ($\dot{S}_{\rm{brc}}=0$) each starting with a cell of initial size $x_0=0.5$, with deterministic ($\xi=0$) single cell growth at a rate $\nu=0.8$. We chose a power law for the division rate $r(x)=x^{\alpha}$ and a transition probability $\Sigma(x|x')=b(x/x')/x'$ depending on the ratio of the daughter to mother volumes through the Gaussian distribution $b = \mathcal{N}(1/2,\sigma_{b}^2)$, so that $\Sigma(\cdot|x') = \mathcal{N}(x'/2,(\sigma_{b} x')^2)$. For three different couples $(\alpha,\sigma_{b})$, the curves exhibit a positive region in the transient dynamics (above the red dashed line), corresponding to a widening of the distribution of sizes due to resetting events from regions of high to low probabilities in size space, and a resulting increase of the Shannon entropy. In the long time limit, they all converge to $-\nu$, independently of $\alpha$ and $\sigma_{b}$.

\subsection{Timer}

The timer mechanism assumes that the division rate is a function of the age of the cell only. It is not described by \cref{eq_FP_1d}, but by a similar equation for which the source term (apparition of new cells) enters as a boundary condition. This is because age is defined on $[0,\infty[$ and the resetting age $a=0$ is on the boundary of the domain. This is the first difference with the case treated in \cite{fuchs_stochastic_2016}, in which the resetting position $x_0$ was inside the domain of definition of $x$. The second difference is that age is by definition the time elapsed since birth and cannot undergo thermal fluctuations, therefore its dynamics between divisions is deterministic, which corresponds to a temperature $T=0$, a mobility $\mu=1$ and a force $F(a)=1$, deriving from a potential $V(a)=-a$.
The signs of the different contributions to the first law are the same as those found for the sizer case. Indeed, because the potential $V(a)$ is decreasing with age, $\dot{W}_{\rm{rst}} \leq 0$, and since $r(a)$ is generally an increasing function of age for non extreme-ages \cite{robert_division_2014}, the covariance between $V$ and $r$ is negative, and so $\dot{W}_{\rm{brc}} \geq 0$. We also find that $\dot{Q}=1 \geq 0$. Finally, the timer is described by:
\begin{eqnarray}
\label{eq_FP_age}
\partial_t p(a) &= -\partial_a p(a) - \lbk \Lambda + r(a) \rbk p(a) \\
\label{eq_FP_age_CI}
p(0) &= 2 \int \di a \ r(a) p(a) \,,
\end{eqnarray}
where the boundary condition accounts for the divisions of cells of any age, giving birth to two new cells of age $a=0$ regardless of the age of the dividing cell.

Integrating eq. (\ref{eq_FP_age}) over $a$ and using the boundary condition leads to the following relation:
\be
\label{eq_p0_lam}
\Lambda= \int \di a \ r(a) p(a) = \frac{p(0)}{2}  \,.
\ee
Following the same steps, we derive the time-evolution of the Shannon entropy of the age distribution:
\begin{eqnarray}
\fl \dot{S}_{\rm{sys}} =  p(0)  - \Lambda S_{\rm{sys}} -  \int \di a \ r(a) p(a) \ln p(a) 
 + 2 \int \di a \ r(a) p(a) \ln \lp \frac{p(a)}{p(0)} \rp \,,
\end{eqnarray}
where we identify
\begin{eqnarray}
\dot{S}_{\rm{brc}} &=- \Lambda S_{\rm{sys}} - \int \di a \ r(a) p(a) \ln p(a) \,, \\
\label{eq_S_brc_cov_timer}
&= - \Cov (r, \ln p)
\end{eqnarray}
and
\be
\label{eq_S_rst_age}
\dot{S}_{\rm{rst}}= 2 \int \di a \ r(a) p(a) \ln \lp \frac{p(a)}{p(0)} \rp \,.
\ee
Finally, the steady-state second law for the timer reads:
\be
\label{eq_2law_timer}
2 \Lambda = -\dot{S}_{\rm{rst}}-\dot{S}_{\rm{brc}} \,,
\ee
where the term $2\Lambda$ does not arise from the force divergence, which is null in this case, but from the boundary condition eq. (\ref{eq_FP_age_CI}). Let us mention that in the case where cells divide in $m$ daughter cells, the factor $2$ in the boundary condition eq. (\ref{eq_FP_age_CI}) is replaced by $m$, \cref{eq_p0_lam} changes to $m \Lambda = (m-1) p(0)$, which results in a coefficient $m/(m-1)$ instead of $2$ for the term $\Lambda$ in the second law \cref{eq_2law_timer}.

One can solve eq. (\ref{eq_FP_age}) for the steady-state age distribution \cite{garcia-garcia_linking_2019}:
\be
\label{eq_pa_ss}
p(a)=p(0) \exp \lbk -\Lambda a - \int_{0}^{a} r(a') \di a' \rbk \,,
\ee
which allows to compute explicitly the resetting and branching entropy production rates:
\begin{eqnarray}
\dot{S}_{\rm{rst}}&= 2 \Lambda \lp S_{\rm{sys}} + \ln \Lambda  + \ln 2 -2  \rp \label{eq_S_rst_lim} \\
\dot{S}_{\rm{brc}}&= 2 \Lambda \lp -S_{\rm{sys}} - \ln \Lambda - \ln 2 +1 \rp \label{eq_S_brc_lim} \,,
\end{eqnarray}
which are functions of $\Lambda$ and $S_{\rm{sys}}$, themselves only depending on the branching rate $r(a)$. 
Moreover, the steady-state age distribution \cref{eq_pa_ss} is a decreasing function of age, which implies that (i) the branching entropy production rate is positive: $\dot{S}_{\rm{brc}} \geq 0$ by eq. (\ref{eq_S_brc_cov_timer}), (ii) the resetting entropy production rate is negative: $\dot{S}_{\rm{rst}} \leq 0$ by \cref{eq_S_rst_age}.

\subsection{Adder}

In the adder theory, the distribution of added volume between birth and death is independent of the volume at birth. This strategy has been found to be in very good agreement with experimental data for many cell types \cite{campos_constant_2014,taheri-araghi_cell-size_2015}. Two variables are required to model the adder, for example the size $x$, which undergoes relative resetting, and the volume added since birth $\Delta = x- x_b$, which undergoes absolute resetting to value $0$, and where $x_b$ is the size at birth. The dynamics of the adder is given by:
\begin{eqnarray}
\label{eq_FP_adder}
\partial_t p(x,\Delta) &= -\lp \partial_x + \partial_{\Delta} \rp \lbk F(x) p(x,\Delta) \rbk - \lbk \Lambda + r(x,\Delta) \rbk p(x, \Delta) \\
\label{eq_FP_adder_CI}
F(x)p(x,0) &= 2 \int \di x' \di \Delta' \ r(x',\Delta') p(x',\Delta') \Sigma(x|x') \,,
\end{eqnarray}
where the boundary condition in the second line accounts for the absolute resetting of the added volume $\Delta$ at division, and can be understood as a balance for any size $x$ between the birth rate of newborn cells of size $x$ (r.h.s.) and the rate at which newborn cells of size $x$ grow (l.h.s.).
Moreover, to obtain the adder property, that is $P(\Delta|x_b)=P(\Delta)$ when growth is deterministic: $\di x/\di t = F(x)$, then the division rate per unit time has to take the form $r(x,\Delta)=F(x) \gamma(\Delta)$, where $\gamma(\Delta)$ is the division rate per unit volume \cite{taheri-araghi_cell-size_2015}.

Conducting the same analysis as before, we derive the steady-state second law in the case of deterministic growth, for the joint distribution $p(x, \Delta)$ of volume and added volume:
\be
\label{eq_2law_adder}
3 \Lambda = -\dot{S}_{\rm{rst}}-\dot{S}_{\rm{brc}} \,.
\ee
The l.h.s. is the sum of the contributions $2\Lambda$ (similar to that of the timer), coming from the boundary condition at $\Delta=0$, and $\Lambda$ (similar to that of the sizer), arising from the force-divergence entropy due to the growth of cells at a rate $\nu$ (equal to $\Lambda$ in steady-state). Note that since the adder is a multivariate model with two variables, one could expect the force-divergence entropy to be a sum of two terms, as detailed in \ref{sec_nd}. However, the force $F$ here is supposed to be a function of the size $x$ only and independent of the added volume $\Delta$, thus $\partial_{\Delta} F(x)=0$.

In the spirit of the last part of \ref{sec_nd}, one can coarse-grain the variable $\Delta$ by integrating eq. (\ref{eq_FP_adder}) on $\Delta$, leading to an equation of the form of a sizer (\cref{eq_FP_1d}) with marginal probability $\hat{p}(x)=\int \di \Delta \ p(x, \Delta)$ and coarse-grained branching rate $\hat{r}(x)  = F(x) \int \di \Delta \ p(\Delta|x) \gamma(\Delta)$. Thus, the steady-state second law for the marginal size distribution obeys \cref{eq_2law_sizer} with the coarse-grained branching rate. 
Without branching, this second law reduces to $\dot{S}_{\rm{rst}}=-\nu$. In the third remark of \cref{sec_sizer} we already noted that the entropy production rate due to resetting was independent of the division rate, we see now that is also independent of the size control mechanism, namely sizer or adder.

\subsection{Analogy with heat engines}

\begin{figure}
	\includegraphics[width=0.49\linewidth]{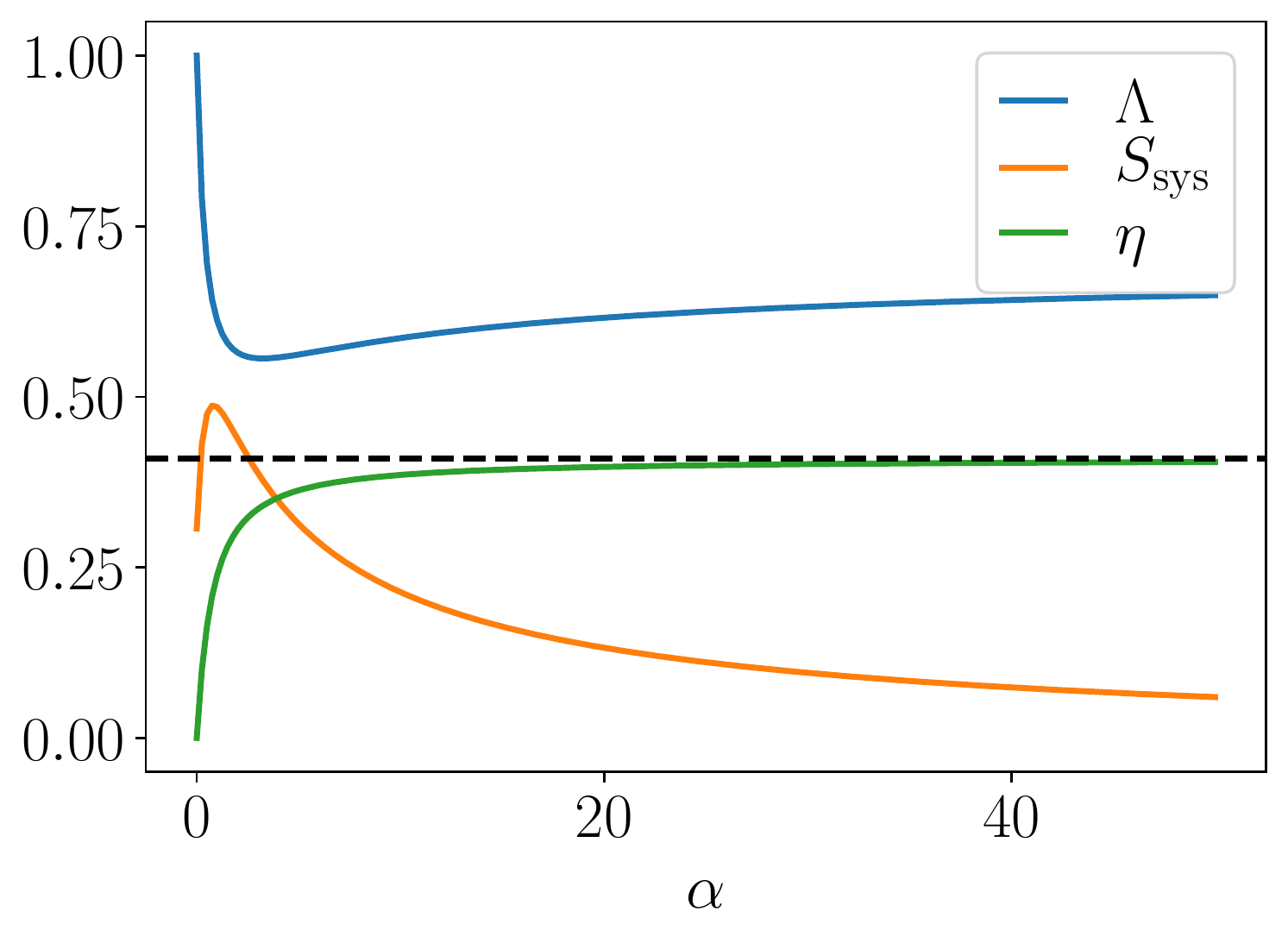}
	\includegraphics[width=0.49\linewidth]{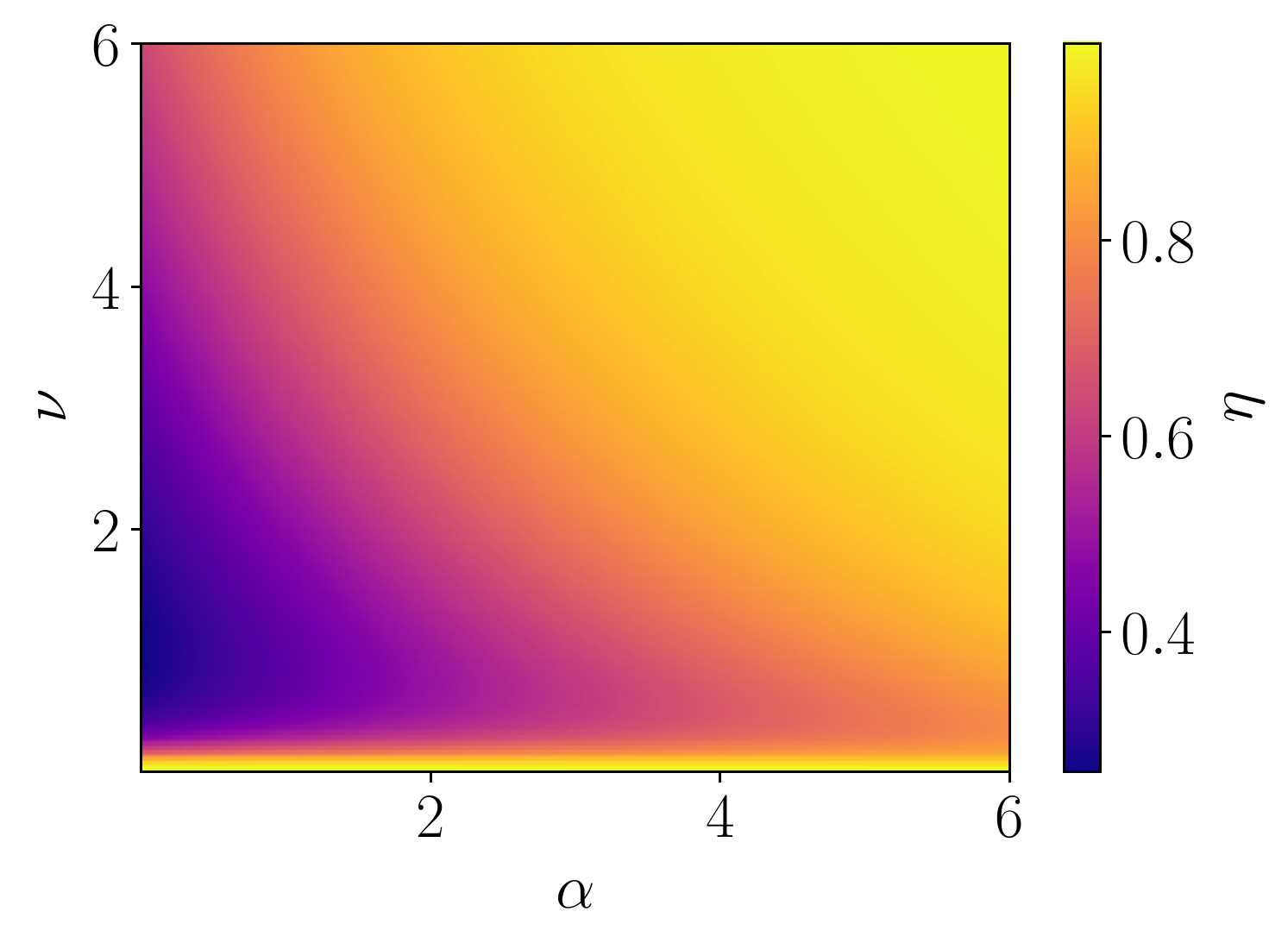}
	\caption{Numerical evaluation of the efficiency $\eta$ when varying parameters of the model for the timer (left) and sizer (right). \textbf{(Left)}: When increasing the strength $\alpha$ of the age control, abnormally old cells disappear and the age distribution narrows around young cells, leading to a decrease in the Shannon entropy (orange). The efficiency (green) monotonously increases with $\alpha$, up to the limit $\ln 2 / (1 + \ln 2)$ (black dashed line). \textbf{(Right)}: For any $\nu$ ($=\Lambda)$, the efficiency is a monotonously increasing function of the size control $\alpha$, up to the maximum value $1$. For any $\alpha$, the efficiency tends to $1$ in both the $\nu \rightarrow 0$ (Carnot) and $\nu \rightarrow \infty$ limits, with a single minimum in between.}
	\label{fig_eff}
\end{figure}

For each of the division-control models studied above, we obtained a second law in the form of $\beta \Lambda = -\dot{S}_{\rm{rst}}-\dot{S}_{\rm{brc}}$ (\cref{eq_2law_sizer,eq_2law_timer,eq_2law_adder}), with $\beta$ an integer equals to $1$ for the sizer, $2$ for the timer, and $3$ for the adder. 
The form of this common second law suggests the definition of the efficiency 
\be
\label{eq_def_eff}
\eta=\frac{-\dot{S}_{\rm{brc}}}{\dot{S}_{\rm{rst}}} \,.
\ee
This definition is inspired by thermodynamic machines, which operate with a driving process and an output process, respectively associated with the entropy production rates $\sigma_1 \geq 0$ and $\sigma_2 \leq 0$. In that case, the thermodynamic efficiency reads $\eta = - \sigma_2/\sigma_1 \geq 0$. 
The second law $\sigma_{tot} = \sigma_1 + \sigma_2 \geq 0$, where $\sigma_{tot}$ is the total entropy production rate, further implies that $\eta \leq 1$. 
In our case, despite the absence of a thermostat and therefore the absence of a first law, by analogy, $\beta \Lambda  \geq 0$, $-{\dot{S}_{\rm{rst}}}$ and $-{\dot{S}_{\rm{brc}}}$ play the roles of $\sigma_{tot}$, $\sigma_1$ and $\sigma_2$, respectively. Indeed, we proved for the sizer and the timer that $-{\dot{S}_{\rm{rst}}}$ and $-{\dot{S}_{\rm{brc}}}$ have the same sign as $\sigma_1$ and $\sigma_2$, respectively.
This analogy in which resetting is the driving process that enables branching, which is the output process, can also be understood intuitively at the level of energies. Populations of cells thrive by dividing, a process for which the creation of a new cell is made possible by the size reduction of both the mother cell and the newborn cell. Indeed, we proved for both the sizer and timer that $\dot{W}_{\rm{rst}} \leq 0$ and $\dot{W}_{\rm{brc}} \geq 0$, which fundamentally comes from the non-confining shapes of the potentials $V(x)$ and $V(a)$. This implies that branching has an energetic cost for the colony, which is covered by the energetic gain due to resetting. 

For the timer, we proved that ${\dot{S}_{\rm{rst}}}$ and ${\dot{S}_{\rm{brc}}}$ are only functions of the branching rate $r(a)$, which we describe by a power law $r(a)=a^{\alpha}$. Thus the strength ${\alpha}$ of the age control is the only parameter in the model.
On \cref{fig_eff} (left), we plot the evolution of the Shannon entropy of the age distribution, the population growth rate and the efficiency against $\alpha$. 
The population growth rate $\Lambda$ is obtained by plugging the steady-state solution \cref{eq_pa_ss} into the boundary condition eq.(\ref{eq_FP_age_CI}) and solving numerically for $\Lambda$. Knowing $\Lambda$, the Shannon entropy of the age distribution is numerically evaluated, and so is the efficiency using eqs. (\ref{eq_S_rst_lim},\ref{eq_S_brc_lim}).
The efficiency is an increasing function of $\alpha$, and converges to the asymptotic value computed in \ref{sec_eff_tim}:
\be
\label{eq_eff_tim}
\lim\limits_{\alpha \rightarrow +\infty} \eta = \frac{\ln 2}{1 + \ln 2} \approx 0.41 \,,
\ee
as the strength of the control increases, leading to a synchronized population where all cells deterministically divide at age $1$.

More parameters are required to describe the sizer: the strength $\alpha$ of the control in the branching rate $r(x)=x^{\alpha}$, and also the single cell growth rate $\nu$ and the parameters of the the kernel $\Sigma$. Here, we focus on symmetric division, thus our model has two parameters: $\alpha$ and $\nu$. We plot on \cref{fig_eff} (right) the efficiency against $\alpha$ ($x$-axis) and $\nu$ ($y$-axis).
We compute the branching entropy production rate using \cref{eq_S_brc_sizer} and the parameters 
$\mu = \ln \nu/\alpha - \ln 2/4$ and $\sigma^2 = \ln 2/2 \alpha$, that can be obtained following the method proposed in \cite{hosoda_origin_2011} as detailed in \ref{sec_lognorm}.
For any $\nu$, the efficiency is an increasing function of $\alpha$, and converges to $1$, the maximal efficiency. This comes from the fact that $\Lambda$ is independent of $\alpha$, and $\dot{S}_{\rm{brc}} \rightarrow \infty$ as $\alpha \rightarrow \infty$ from \cref{eq_S_brc_sizer}. 
For any $\alpha$, the efficiency starts at $1$ when $\nu=0$ (analogous to Carnot efficiency when the entropy production is null), decreases until reaching a minimum and then tends to $1$ (obtained from the rate of increase of $\dot{S}_{\rm{brc}}$ with $\nu$) as $\nu$ is varied from $0$ to $\infty$.

Note that the Shannon entropy decreases as the strength of control measured by $\alpha$ is increased in the region where $\alpha$ is large. This means that the diversity in the controlled trait is reduced across the population. We suspect that such a lack of diversity might be harmful for the population at some level, which could be one of the reasons why cells need not implement such a strong and efficient (in the sense of $\eta$) control.

We remind the reader that the two plots cannot be compared directly, in order to find the most efficient control strategy for example, since the efficiencies plotted are related to different distributions. Indeed, the efficiency for the timer is the ratio of the branching to resetting entropy production rates associated with the age distribution, versus the size distribution for the sizer.

\section{Conclusion}

The growth and division of cells within a colony is clearly an irreversible process, which must be  
constrained by the laws of thermodynamics. One way to approach this question is to study the corresponding 
breaking of the time reversal symmetry in that process \cite{england_statistical_2013}. 
Another way, which we pursue here, is
to describe cell growth and division in terms of two subprocesses, branching and resetting, 
which can then be analyzed separately using Stochastic Thermodynamics. 

We find from such an approach that resetting is a process which rises the internal energy of the system, thereby allowing the system 
to pay the energy cost associated with branching. Branching is required by cells to self-replicate, and in doing so, to maintain 
certain traits within lineages, an essential feature for the survival of the cell colony.
In exponentially growing colonies, the population growth rate emerges from a carefully controlled balance between resetting and branching. 
We introduce an efficiency, akin to the thermodynamic efficiency of thermal machines, which quantifies 
the energy conversion between these two processes.

The present work could be extended in several future directions. 
The assumption that the colony is in an exponentially growing phase could be relaxed, and  
additional sources of stochasticity affecting single-cell growth rate could be included \cite{thomas_sources_2018}.
To improve the description of the adder model, it would be interesting to obtain analytical solutions 
for simple choices of the resetting rate.
Recently, several experimental works have suggested that the adder model can be justified microscopically using incremental threshold models \cite{pandey_exponential_2020}. Along the same line, some recent works inspired by \cite{nieto_unification_2020} proposed a unified model for cell size distributions accounting for the sizer, timer and adder behaviors, using an $N$-stages description of the cell cycle, which is applicable to bacterial exponential growth \cite{jia_cell_2021} and to yeast growth \cite{jia_characterizing_2021}.
We note that the incremental model and the $N$-stages model both fall into the class of resetting models studied here, and for that reason it would be interesting to adapt our approach to these more recent models. 
In addition, the present framework is not limited to models of cell size control, it is also relevant to describe the homeostasis of other variables distinct from the cell size or age. 

In the three mechanisms of cell size control we studied, the opposite of the sum of the resetting and branching entropy production rates
equals the population growth rate times a prefactor which seems to depend on the number of key variables of the model. Indeed, 
this factor is one for the sizer, two for the timer and three for the adder.
In principle, the prefactor could take other values depending on the number of variables entering in the growth function. 
Interestingly, this prefactor might therefore give insight into a particular mechanism of control and 
possibly be related to the latent variables, 
which are detected in Bayesian approaches of lineage data \cite{nakashima_lineage_2020}. 

\ack
The authors warmly thank Jeremie Unterberger for a critical reading and stimulating remarks on the present work.

\appendix

\section{Multi-dimensional systems}
\label{sec_nd}

For simplicity, in the main text we presented our results in one dimension, but we show in this appendix that they hold in $n$ dimensions. Let $\xbf=(x_1, ..., x_n)$ be a $n$-dimensional vector, and let the branching rate $r(\xbf)$ and the resetting kernel $\Sigma(\xbf|\bi{x'})$ depend on these $n$ variables. For more generality, and regarding the discussion of \cref{sec_T_0}, we consider that $k$ variables $(x_1, ..., x_k)$ are in contact with different heat baths $\{ T_i\}_{i=1,...,k}$, and that $n-k$ variables $(x_{k+1}, ..., x_n)$ are not linked to any heat bath and thus do not undergo thermal fluctuations. Among those $n-k$ athermal degrees of freedom, some are subjected to deterministic forces $F$, while others stay constant between resetting events ($F=0$). We label $(x_{k+1}, ..., x_{k+l})$ the $l$ variables for which $F \neq 0$, and $(x_{k+l+1}, ..., x_{n})$ the $n-k-l$ variables for which $F=0$. In this setting, the potential $V(\xbf)$ only depends on the $k+l$ first variables, which defines the same number of forces: $F_i(\xbf)=-\partial_{x_i} V(\xbf)$. For these degrees of freedom, we define the mobilities $\mu_i$ and the currents $j_i(\xbf)=\mu_i F_i(\xbf) p(\xbf) - \mu_i T_i \partial_{x_i} p(\xbf)$ for $i=1, ..., k$, and $j_i(\xbf)=\mu_i F_i(\xbf) p(\xbf)$ for $i=k+1, ..., k+l$, and recall that for the variables $(x_{k+l+1}, ..., x_{n})$ there is current. For clarity, we define the vectors of currents $\bi{j}(\xbf)$, forces $\bi{F}(\xbf)$, and mobilities $\boldsymbol{\mu}$, whose components are respectively $j_i(\xbf)$, $F_i(\xbf)$ and $\mu_i$ for $i=1, ..., k+l$. 

Finally the equation for the evolution of $p(\xbf)$ reads:
\be
\label{eq_FP_nd}
\partial_t p(\xbf) = - \bi{\nabla} \cdot \bi{j}(\xbf) - \lbk \Lambda + r(\xbf) \rbk p(\xbf) + m \int \di \xbf' \Sigma(\xbf|\xbf') r(\xbf') p(\xbf') \,,
\ee

The first law of thermodynamics is obtained following the same steps as in the 1d case, and when replacing $x$ by $\xbf$, the expressions of $\dot{W}_{\rm{rst}}$ (\ref{eq_W_rst}) and $\dot{W}_{\rm{brc}}$ (\ref{eq_W_brc_cov}) are unchanged. The heat is replaced by a sum of heats associated with each degree of freedom:
\begin{eqnarray}
\dot{Q} &= \int \di \xbf \ \bi{j}(\xbf) \cdot \bi{F}(\xbf) \\
& = \sum_{i=1}^{k} \int \di \xbf \ j_i(\xbf) F_i(\xbf) + \sum_{i=k+1}^{k+l} \int \di \xbf \ \mu_i  p(\xbf) F_i^2(\xbf)\,,
\end{eqnarray}
where for $i=1, ..., k$, $\dot{Q}_i=\int \di \xbf \ j_i(\xbf) F_i(\xbf)$ is the heat exchange rate with the $i^{\rm{th}}$ thermostat, and for $i=k+1, ..., k+l$, $\mu_i \langle F_i^2 \rangle$ is the rate of change of the athermal heat discussed in \cref{sec_T_0}, associated with the $i^{\rm{th}}$ degree of freedom.

The second law also follows from the same calculation and the entropy production rates due to branching (eq. (\ref{eq_S_brc_cov})) and resetting (\cref{eq_S_rst}) are unchanged as compared to the 1d case. 
The term due to currents can be split into two contributions corresponding to the thermal and athermal degrees of freedom:
\begin{eqnarray}
\fl & - \int \di \xbf \ \frac{\bi{\nabla}(p(\xbf)) \cdot \bi{j}(\xbf)}{p(\xbf)} \nn \\
\fl &= - \int \di \xbf \ \lbk \sum_{i=1}^{k} \frac{\lp F_i(\xbf)p(\xbf)- \mu_i^{-1} j_i(\xbf) \rp j_i(\xbf)}{T_i p(\xbf)} - \sum_{i=k+1}^{k+l} \mu_i F_i(\xbf) \partial_{x_i} p(\xbf)  \rbk \nn \\
\fl &= \sum_{i=1}^{k} \lbk \int \di \xbf \lp \frac{j_i^2(\xbf)}{\mu_i T_i p(\xbf)} - \frac{ F_i(\xbf) j_i(\xbf)}{T_i } \rp \rbk + \sum_{i=k+1}^{k+l} \mu_i \int \di \xbf \ p(\xbf) \partial_{x_i} F_i(\xbf) \,,
\end{eqnarray}
where $\dot{S}_{\rm{c}}^{(i)}=\int \di \xbf \ j_i^2(\xbf)/ \mu_i T_i p(\xbf) \geq 0$ is the entropy production rate due to the non-equilibrium current associated with the degree of freedom $i$, $\dot{S}_{\rm{m}}^{(i)}=\int \di \xbf \  F_i(\xbf) j_i(\xbf)/T_i = \dot{Q}_i/T_i$ is the entropy exchange rate with thermostat $i$, and $\dot{S}_{\rm{fd}}^{(i)}=\mu_i \int \di \xbf \ p(\xbf) \partial_{x_i} F_i(\xbf)$ is the force-divergence entropy production rate associated with the athermal degree of freedom $i$.
Finally the second law for $n$-dimensional systems reads:
\be
\dot{S}_{\rm{sys}}=\dot{S}_{\rm{brc}}+\dot{S}_{\rm{rst}}+\sum_{i=1}^{k} \lp  \dot{S}_{\rm{c}}^{(i)} - \dot{S}_{\rm{m}}^{(i)} \rp + \sum_{i=k+1}^{k+l} \dot{S}_{\rm{fd}}^{(i)}\,,
\ee
Note that with the unification of deterministic and entropic forces, we obtain again \cref{eq_2nd_law_v2-1}, now with the identification
\be
\label{entropic-multivariate}
\dot{\tilde{S}}_{\rm{fd}}=-\sum_{i=1}^k \mu_iT_i\langle\partial^2_{x_i}\ln p\rangle+\sum_{i=1}^{k+l}\mu_i\langle\partial_{x_i}F_i\rangle\,,
\ee
ecompassing the entropic contribution of the degrees of freedom which are in contact with a thermal reservoir. We note that the contribution of the deterministic forces (last term in the r.h.s. of \cref{entropic-multivariate}) takes the form $\sum_{i=1}^{k+l}\mu_i\langle\partial_{x_i}F_i\rangle= \langle \rm{div} (\boldsymbol{\mu} \circ \bi{F}) \rangle$, which is the average value of the divergence of the Hadamard product $\boldsymbol{\mu} \circ \bi{F}$, defined by the components $(\boldsymbol{\mu} \circ \bi{F})_i= \boldsymbol{\mu}_i \bi{F}_i$. 

The multivariate model is particularly interesting when the dynamics of resetting and branching is controlled by a set of hidden variables, while we have access to only one observable, let us say $x_1$. In this case, the natural quantity to look at is the Shannon entropy of the marginal distribution of $x_1$: $\hat{p}(x_1)=\int \di x_2 ... \di x_n \ p(x_1,...,x_n)$. In this case, we show that the computation of $\dot{S}_{\rm{brc}}$ and $\dot{S}_{\rm{rst}}$ reduces to a 1d problem with coarse-grained branching rate and resetting kernel.
To obtain the evolution of the Shannon entropy associated with $\hat{p}(x_1)$, we integrate \cref{eq_FP_nd} over $x_2, ..., x_n$, multiply it by $\ln \hat{p}(x_1)$ and integrate it over $x_1$. The contribution of resetting it given by:
\begin{eqnarray}
\dot{S}_{\rm{rst}}& =m \int \di \xbf \ r(\xbf) p(\xbf) \lbk \ln \hat{p}(x_1) - \langle \ln \hat{p} \rangle_{\rho_{nb}} \rbk \\
&=m \int \di x_1 \ \hat{r}(x_1) \hat{p}(x_1) \lbk \ln \hat{p}(x_1) - \langle \ln \hat{p} \rangle_{\hat{\rho}_{nb}} \rbk \,,
\end{eqnarray}
where we defined the coarse-grained branching rate
\be
\hat{r}(x_1) =\int \di x_2 ... \di x_n \ p(x_2,...,x_n|x_1) r(x_1, ..., x_n) \,,
\ee
and where the newborn distribution is also marginalized:
\be
\hat{\rho}_{nb}(x_1)=\frac{\int \di \xbf' \ \hat{\Sigma}(x_1|\xbf') r(\xbf') p(\xbf')}{\int \di x_1 \ \hat{r}(x_1) \hat{p}(x_1)} \,,
\ee
with the coarse-grained kernel $\hat{\Sigma}(x_1|\xbf')=\int \di x_2 ... \di x_n \  \Sigma(x_1, ..., x_n|\xbf')$.
Similarly, the branching entropy production rate reduces to
\begin{equation}
\dot{S}_{\rm{brc}}=-(m-1) \Cov_{\hat{p}} (\hat{r}, \ln \hat{p}) \,.
\end{equation}

Before closing the present discussion on multi-dimensional systems,  it is worth mentioning a subtlety of the coarse-grained description introduced above. Note that when reducing a multivariate problem by eliminating degrees of freedom, the dynamics associated to the remaining variables is non-Markovian in general. Thus, as far one is concerned with one-time observables, the above description is correct because the Markovian dynamics that is obtained when integrating out the unobserved degrees of freedom corresponds to a \emph{substitute} Markov process having the same one-time statistics as the underlying marginal non-Markov dynamics~\cite{Hanggi1977,HANGGI1982207,RGG2012Nonadiabatic}. However, multi-time observables (e.g., correlation functions) are not well captured by such effective Markov dynamics unless some sort of Markov approximation makes sense. In particular, using the path integral representation of the substitute process to compute generating functionals of the marginal non-Markovian degrees of freedom would be misleading, typically yielding incorrect results.

\section{Sign of the branching entropy production rate for the sizer in steady-state}
\label{app_S_brc}

In this section, we show that if the steady state size distribution is log-normal:
\be
p(x)= \frac{1}{x \sigma \sqrt{2 \pi}} \exp \lbk \frac{-(\ln x - \mu)^2}{2 \sigma^2} \rbk \,,
\ee
with parameters $\mu$, $\sigma > 0$, and that the division rate is a power law $r(x)= x^{\alpha}$, where $\alpha \geq 0$ is the strength of the size control, then the entropy production rate due to branching, given by eq. (\ref{eq_S_brc_cov}), is positive. To do so, one needs to compute the covariance between the division rate and the logarithm of the size distribution:
\be
\label{eq_def_cov}
\Cov(r,\ln p)=\langle r \ln p \rangle - \langle r \rangle \langle \ln p \rangle \,,
\ee
where the moments of the log-normal distribution are known:
\be
\label{eq_logn_moments}
\langle x^{\alpha} \rangle = \exp \lbk \alpha \mu + \frac{\alpha^2 \mu^2}{2} \rbk \,,
\ee
as well as its entropy:
\be
\label{eq_logn_entropy}
\langle \ln p \rangle = -\frac{1}{2} - \mu - \ln (\sigma \sqrt{2 \pi}) \,.
\ee
Therefore, one only needs to compute the term:
\begin{eqnarray}
\fl \langle r \ln p \rangle & =  - \int_{0}^{\infty} \di x \ \frac{x^{\alpha}}{x \sigma \sqrt{2 \pi}} \lbk (\ln x - \mu)^2 + \ln x + \ln (\sigma \sqrt{2 \pi}) \rbk \exp \lbk \frac{-(\ln x - \mu)^2}{2 \sigma^2} \rbk \nn \\
\fl & = \frac{e^{\frac{-\mu^2}{2 \sigma^2}}}{\sigma \sqrt{2 \pi}} \int\limits_{-\infty}^{+ \infty} \!  \di u \! \lbk \frac{-u^2}{2 \sigma ^2} \! + \! u \lp \frac{\mu}{\sigma^2} \! - \! 1 \rp \rbk e^{\frac{-u^2}{2 \sigma^2} + u \lp \frac{\mu}{\sigma^2} + \alpha \rp } - \lp \ln (\sigma \sqrt{2 \pi}) \! + \! \frac{\mu^2}{2 \sigma^2} \rp \langle x^{\alpha} \rangle  \nn \\
\label{eq_r_lnp}
\fl & = - \langle x^{\alpha} \rangle  \lbk  \ln (\sigma \sqrt{2 \pi}) + \mu + \frac{1}{2} + \alpha \sigma^2 + \frac{\alpha^2 \sigma^2}{2}  \rbk \,,
\end{eqnarray}
where the second line is obtained with the change of variable $u=\ln x$, and the third line comes from the resolution of the integral in the form of a Gaussian function times a second order polynomial, which is analytical. 

Finally, combining eqs. (\ref{eq_S_brc_cov}) and \ref{eq_def_cov} to \ref{eq_r_lnp} leads to
\be
\dot{S}_{\rm{brc}} = (m-1) \alpha \sigma^2 \lp 1 + \frac{\alpha}{2} \rp \exp \lbk \alpha \mu +  \frac{\alpha^2 \mu^2}{2} \rbk \geq 0 \,,
\ee
which is positive regardless of the actual values of $\mu$, $\sigma$ and $\alpha$.

\section{Asymptotic efficiency for the timer}
\label{sec_eff_tim}

In this section, we demonstrate \cref{eq_eff_tim} giving the value of the timer efficiency in the limit of strong age-control ($\alpha \rightarrow \infty$). In this limit, the power-law branching rate $r(a)=a^{\alpha}$ becomes a step, taking value $0$ between $0$ and $1$ and diverging for $a > 1$, therefore cells deterministically divide at age $1$. Consequently, all divisions are synchronized in the population and $N(t)=N_0 2^{\lfloor t \rfloor}$ where $\lfloor t \rfloor$ is the integer part of $t$. Moreover, in the long time limit, the long term population growth rate $\Lambda_t=1/t \int_{0}^{t} \di t' \Lambda(t')=1/t \ln \lp N(t)/N_0 \rp$ is equal to the instantaneous growth rate $\Lambda$.
Thus,
\begin{eqnarray}
\label{eq_lambda_lim}
\Lambda&= \lim\limits_{t \rightarrow \infty} \frac{\lfloor t \rfloor}{t} \ln 2 \nn \\
& = \ln 2 \,.
\end{eqnarray}
Using this result, we can compute the Shannon entropy of the steady-state age distribution (\cref{eq_pa_ss}):
\begin{eqnarray}
\fl S_{\rm{sys}} &=- \int_{0}^{\infty} \di a \ p(a) \ln p(a) \nn \\
\fl &=- \ln (2 \Lambda) \int_{0}^{\infty} \di a \ p(a) + 2 \Lambda \int_{0}^{\infty} \di a \lp \Lambda a + \int_{0}^{a} \di a' \ r(a') \rp \exp \lbk - \Lambda a - \int_{0}^{a} \di a' \ r(a') \rbk \nn \\
\fl & =- \ln (2 \Lambda) + 2 \Lambda \int_{0}^{1} \di a \ \Lambda a \ \exp \lbk - \Lambda a \rbk \nn \\
\fl &= 1- 2 \ln 2 - \ln(\ln 2) \,,
\label{eq_S_lim}
\end{eqnarray}
We went from line $2$ to line $3$ using the normalization of $p$ for the first integral, and we decomposed the second integral into two parts: from $0$ to $1$ and from $1$ to $\infty$. Between $1$ and $\infty$, the contribution of the branching rate diverges and thus the integral is nullified by the exponential function, and only the integral from $0$ to $1$ remains, for which $r(a) \approx 0$.

Finally, we plug eq. (\ref{eq_lambda_lim}) and eq. (\ref{eq_S_lim}) into  eq. (\ref{eq_S_rst_lim}) and eq. (\ref{eq_S_brc_lim}), and plug them into the definition of the efficiency \cref{eq_def_eff} to obtain:
\be
\eta=\frac{\ln 2}{1 + \ln 2} \approx 0.41 \,.
\ee

\section{Parameters of the log-normal steady-state size distribution}
\label{sec_lognorm}

We follow in this section the analysis conducted in \cite{hosoda_origin_2011} to determine the parameters $(\mu, \sigma)$ of the log-normal ansatz for the steady-state size distribution. 

We multiply \cref{eq_FP_1d}, in the case where $F(x)=\nu x$, $\mu=1$ and $D=0$ which describes the sizer with deterministic growth, by $x^k$ and integrate over $x$ to recast the steady-state equation as an equation for the moments of $p(x)$:
\be
\label{eq_FP_moments}
\nu k \langle x^k \rangle - \langle x^{k+\alpha} \rangle - \langle x^{k} \rangle \langle x^{\alpha} \rangle + 2 \int \di y \ y^{\alpha}  p(y) \int \di x \ x^k \Sigma(x|y) = 0\,.
\ee
We consider a kernel $\Sigma(x|y)=b(x/y)/y$ with $b = \mathcal{N}(1/2,\sigma_{b}^2)$ so that the transition probability is Gaussian and centered around symmetric division: $\Sigma(\cdot|y) = \mathcal{N}(y/2,(y \sigma_{b})^2)$. The last integral in the above expression gives the moment of the Gaussian distribution, which are known. 

Experimental data are well accounted for by log-normal distributions, so we plug the ansatz $p = \rm{Lognormal} (\mu ,\sigma^{2})$ in \cref{eq_FP_moments}, where the moments of $p$ are given by \cref{eq_logn_moments}. 
Then, evaluating the expression for $k=1$ and $k=2$ gives:
\begin{eqnarray}
\mu &= \frac{\ln \nu}{\alpha} - \frac{1}{4} \ln \lp \frac{2}{1-4\sigma_{b}^2} \rp \\
\sigma^2 &= \frac{1}{2 \alpha} \ln \lp \frac{2}{1-4\sigma_{b}^2} \rp \,.
\end{eqnarray}
Even though data are often fitted by the log-normal distribution with the parameters derived above, it is not a solution to \cref{eq_FP_1d} in steady-state, and becomes less and less accurate as kernel $\Sigma$ broadens. Therefore, these parameters are valid for symmetric division ($\sigma_{b}=0$), and for near-symmetric division ($\sigma_{b} \ll 1/2$).

\section*{References}

\providecommand{\newblock}{}


\begin{thebibliography}{10}
	\expandafter\ifx\csname url\endcsname\relax
	\def\url#1{{\tt #1}}\fi
	\expandafter\ifx\csname urlprefix\endcsname\relax\def\urlprefix{URL }\fi
	\providecommand{\eprint}[2][]{\url{#2}}
	%
	%
	
	\bibitem{evans_diffusion_2011}
	Evans M~R and Majumdar S~N 2011 {\em Phys. Rev. Lett.\/} {\bf 106} 160601
	
	\bibitem{coppey_kinetics_2004}
	Coppey M, B\'{e}nichou O, Voituriez R and Moreau M 2004 {\em Biophys. J.\/}
	{\bf 87} 1640--1649
	
	\bibitem{benichou_optimal_2005}
	B\'{e}nichou O, Coppey M, Moreau M, Suet P~H and Voituriez R 2005 {\em Phys.
		Rev. Lett.\/} {\bf 94} 198101
	
	\bibitem{evans_stochastic_2020}
	Evans M~R, Majumdar S~N and Schehr G 2020 {\em J. Phys. A: Math. Theor.\/} {\bf
		53} 193001
	
	\bibitem{fuchs_stochastic_2016}
	Fuchs J, Goldt S and Seifert U 2016 {\em EPL\/} {\bf 113} 60009
	
	\bibitem{roldan_path-integral_2017}
	Rold{\'a}n E and Gupta S 2017 {\em Phys. Rev. E\/} {\bf 96} 022130
	
	\bibitem{pal_integral_2017}
	Pal A and Rahav S 2017 {\em Phys. Rev. E\/} {\bf 96} 062135
	
	\bibitem{gupta_work_2020}
	Gupta D, Plata C~A and Pal A 2020 {\em Phys. Rev. Lett.\/} {\bf 124} 110608
	
	\bibitem{pal_thermodynamic_2021}
	Pal A, Reuveni S and Rahav S 2021 {\em Phys. Rev. Research\/} {\bf 3} 013273
	
	\bibitem{eliazar_branching_2017}
	Eliazar I 2017 {\em EPL\/} {\bf 120} 60008
	
	\bibitem{pal_first_2019}
	Pal A, Eliazar I and Reuveni S 2019 {\em Phys. Rev. Lett.\/} {\bf 122} 020602
	
	\bibitem{seifert_stochastic_2012}
	Seifert U 2012 {\em Rep. Prog. Phys.\/} {\bf 75} 126001
	
	\bibitem{garcia-garcia_linking_2019}
	Garc\'{\i}a-Garc\'{\i}a R, Genthon A and Lacoste D 2019 {\em Phys. Rev. E\/}
	{\bf 99} 042413
	
	\bibitem{robert_division_2014}
	Robert L, Hoffmann M, Krell N, Aymerich S, Robert J and Doumic M 2014 {\em BMC
		Biol.\/} {\bf 12} 17
	
	\bibitem{jun_fundamental_2018}
	Jun S, Si F, Pugatch R and Scott M 2018 {\em Rep. Prog. Phys.\/} {\bf 81}
	056601
	
	\bibitem{ho_modeling_2018}
	Ho P~Y, Lin J and Amir A 2018 {\em Annu. Rev. Biophys.\/} {\bf 47} 251--271
	
	\bibitem{jia_cell_2021}
	Jia C, Singh A and Grima R 2021 {\em iScience\/} {\bf 24} 102220
	
	\bibitem{lin_effects_2017}
	Lin J and Amir A 2017 {\em Cell Syst.\/} {\bf 5} 358--367.e4
	
	\bibitem{wang_robust_2010}
	Wang P, Robert L, Pelletier J, Dang W~L, Taddei F, Wright A and Jun S 2010 {\em
		Curr. Biol.\/} {\bf 20} 1099--1103
	
	\bibitem{hosoda_origin_2011}
	Hosoda K, Matsuura T, Suzuki H and Yomo T 2011 {\em Phys. Rev. E\/} {\bf 83}
	031118
	
	\bibitem{osella_concerted_2014}
	Osella M, Nugent E and Cosentino~Lagomarsino M 2014 {\em Proc. Natl. Acad. Sci.
		U.S.A.\/} {\bf 111} 3431--3435
	
	\bibitem{nieto_unification_2020}
	Nieto C, Arias-Castro J, S\'{a}nchez C, Vargas-Garc\'{i}a C and Pedraza J~M
	2020 {\em Phys. Rev. E\/} {\bf 101} 022401
	
	\bibitem{campos_constant_2014}
	Campos M, Surovtsev I~V, Kato S, Paintdakhi A, Beltran B, Ebmeier S~E and
	Jacobs-Wagner C 2014 {\em Cell\/} {\bf 159} 1433--1446
	
	\bibitem{taheri-araghi_cell-size_2015}
	Taheri-Araghi S, Bradde S, Sauls J~T, Hill N~S, Levin P~A, Paulsson J,
	Vergassola M and Jun S 2015 {\em Curr. Biol.\/} {\bf 25} 385--391
	
	\bibitem{england_statistical_2013}
	England J~L 2013 {\em J. Chem. Phys.\/} {\bf 139} 121923
	
	\bibitem{thomas_sources_2018}
	Thomas P, Terradot G, Danos V and Weiße A~Y 2018 {\em Nat. Commun.\/} {\bf 9}
	4528
	
	\bibitem{pandey_exponential_2020}
	Pandey P~P, Singh H and Jain S 2020 {\em Phys. Rev. E\/} {\bf 101} 062406
	
	\bibitem{jia_characterizing_2021}
	Jia C, Singh A and Grima R 2021 {\em bioRxiv:10.1101/2021.06.10.447927\/}
	
	\bibitem{nakashima_lineage_2020}
	Nakashima S, Sughiyama Y and Kobayashi T~J 2020 {\em Bioinformatics\/} {\bf 36}
	2829--2838
	
	\bibitem{Hanggi1977}
	H{\"a}nggi P and Thomas H 1977 {\em Z. Phys., B Condens. matter\/} {\bf 26}
	85--92
	
	\bibitem{HANGGI1982207}
	H{\"a}nggi P and Thomas H 1982 {\em Phys. Rep.\/} {\bf 88} 207--319
	
	\bibitem{RGG2012Nonadiabatic}
	Garc\'{\i}a-Garc\'{\i}a R 2012 {\em Phys. Rev. E\/} {\bf 86} 031117
	
\end{thebibliography}
\end{document}